\RequirePackage{fix-cm}
\documentclass[twocolumn,epjc3]{svjour3}  
\smartqed  
\RequirePackage{graphicx}
\usepackage{amssymb}   
\usepackage[usenames]{xcolor}
\usepackage{amsmath}

\begin{document}

\title{Azimuthal polarization of polarized top quark in noncommutative standard model
}
\author{Z. Rezaei        \and        R. Salehi
}
\thankstext{e1}{e-mail: zahra.rezaei@yazd.ac.ir}

\institute{Yazd University \label{addr1}
}

\maketitle
\begin{abstract}
Top quark decays before hadronization and its information transfer to final products. The top quark decays via the weak interaction, hence the azimuthal polarized decay rate is zero in the leading order of SM.  Therefore the study of top quark could be taken into consideration as useful tools to search for new physics. In this paper, we calculate the decay of polarized top quark $t( \uparrow ) \to b{W^ + } \to b{\ell ^ + }{\upsilon _\ell }$, in the framework of the noncommutative standard model. We investigate the contribution of  noncommutative space-time on the azimuthal polarized decay rate and spin analysing power. Also we depict the energy spectrum for the charged lepton and the variation of the total  differential decay rate with respect to the azimuthal angle $\phi$ which receive some correction in noncommutative space-time.
\end{abstract}

\section{Introduction}\label{section.R1}
Top quark produces  as single top and $\bar{t}t$ pair at tevatron and LHC. It provides a suitable situation to study the top quark properties. 
The top quark is the heaviest known elementary particle having the mass $173.21 \pm 0.51 \pm 0.71\,GeV$ \cite{pdg}. The large mass leads to a very short lifetime for  top quark\cite{n.mah}, in the standard model (SM). The top quark decays via electro-weak interaction before hadronization because its lifetime is an order of magnitude smaller than typical (QCD) hadronization time scale $({\kern 1pt} \frac{1}{{{\Lambda _{QCD}}}} \sim 30{\kern 1pt} \, \times {10^{ - 25}}{\kern 1pt} )$ \cite{n.mah}. Therefore study of top quark could help to probe  the electroweak symmetry breaking mechanism. Also it provides the appropriate circumstance to search for any deviation from the Standard Model and to test the New physics beyond the Standard Model\cite{Aguilar-Saavedra:2017nik}. Using the productions of single top decay and  $\bar{t}t$ pair annihilation, one could obtains some information about the polarization and spin properties of top quark. 
The decay before hadronization cause the spin and the polarization properties transfer to the final products. By studying the polarization properties of the products, the polarization and spin information of the primary particle could be determined.  
When the primary particle's beams that produce the top quarks are polarized, some information could be obtained from them. In hadron colliders, this opportunity don't exist, because the polarization of the primary beams are complicated. But this is possible for future lepton colliders as $e^- e^+$ nad ${\mu}^- {\mu}^+$. We could obtain information from the initial beams as the final products. There are many studies in this subject in hadron colliders \cite{Mahlon:1996pn}
and also for $e^- e^+$ colliders\cite{Kuhn:1985ps,Parke:1996pr,Brandenburg:1998xw}.\\
Top quark propertices are interested from various points of view, the new decay modes, the flavour changing neutral current, the anomalous coupling in the twb vertex, the helicity of $W$ boson, the top quark polarization and the spin correlations  \cite{AguilarSaavedra:2006fy,Fischer:2001gp,Do:2002ky,Fuyuto:2015gmk}. We discuss the single top quark polarization in this paper.\\
 In SM, the top quark decays almost completely into a b-quark and a W-boson with the branching ratio $BR(t \to b{W^ + }) \sim 0.998$ \cite{pdg}. According to the possible leptonic and hadronic decay modes of the ${W^ + }$-boson \cite{a.k}, the dominant semi-leptonic decays of the top quark are $t\to b{W^ + } \to {\kern 1pt} {\kern 1pt} b({\ell ^ + }{\upsilon _\ell })$ in the SM \cite{a.a}. 
There are many studies on the top quark decay properties in SM. We summarize some of them in the following:

- The spin-dependent energy distribution of B-hadrons from polarized top decays is studied by considering the azimuthal correlation rate \cite{dmn3}.

- The QED and QCD radiative corrections to the charged lepton energy distributions are calculated in the dominant semi-leptonic decays of the unpolarized top quark $t\to b{W^ + } \to {\kern 1pt} {\kern 1pt} b({\ell ^ + }{\upsilon _\ell })$ in the SM. The QCD corrections are calculated in the leading and next-to-leading logarithmic approximations, but the QED is considered in the leading logarithmic approximation only \cite{a.a,Bernreuther:2014dla} .

-  $\Gamma _A^0$, $\Gamma _B^0$ and $\Gamma _C^0$ denoting the unpolarized decay rate(UPDR), polar polarized decay rate (PPDR) and azimuthal polarized decay rate(APDR), that are the polarization observables, respectively, are studied for the polarized top quark in the semi-leptonic decay, $t( \uparrow ) \to {X_b} + {\ell ^ + } + {\upsilon _\ell }$. $\Gamma _C^0$ vanishes due to the left chiral $(V-A)$ nature of the weak currents  \cite{Korner:1998nc}.
There are some other examples have been caused from this effect(see\cite{Korner:1998nc} and references there in). Therefore in the leading order (LO) of SM, the APDR is zero. 
The NLO and NLO QCD corrections had been calculated for longitudinal polarized top quark\cite{Groote:1995yc,Do:2002ky}. The NLO QCD corrections is carried out to the $\Gamma _C^0$ for ${m_b} = 0$ and the results is compared to the corresponding contribution of a non-standard-model(N-SM), right-chiral quark current, for ${m_b} \ne 0$. For ${m_b} \ne 0$, it is possible to have the interference contribution from the interference of the left and right chiral quark currents when squaring the full matrix element whereas it is not possible when ${m_b} = 0$. NLO and N-SM right-chiral corrections are $\left| {\Gamma _C^{(1)}} \right| = 2.4 \times {10^{ - 3}}{\kern 1pt} \Gamma _A^0$  and $\left| {\Gamma _C^{(R)} + \Gamma _C^{({\mathop{\rm int}} )}} \right| \le 4.7 \times {10^{ - 5}}{\kern 1pt} \Gamma _A^0$ , respectively \cite{base,diss} where $ {\Gamma _C^{(1)}} $ indicates the APDR, caused by NLO QCD corrections,  ${\Gamma _C^{(R)}}$ is the right chiral quark current contribution and $\Gamma _C^{({\mathop{\rm int}} )}$ display the interference between the SM and right chiral contributions. The absence of the rate $\Gamma _C^0$ in the LO of the SM is a consequence of the left-chiral (V-A)(V-A) nature of the current-current interaction in the SM \cite{base}.

As it is concluded from\cite{base,diss}, nonzero contributions to the rate $\Gamma _C^0$ can either arise from N-SM effects or from higher order SM radiative corrections. Many studies is done on the top quark features via N-SM effects. The nonstandard effects on the full top width have been investigated in the minimal super-symmetric standard model and in the technicolor model \cite{3n1,3n2,3n3}. Also the effects  of  anomalous  tWb  couplings is investigated  on  the top width and  some constraints have been applied on the anomalous couplings \cite{4n1,4n2,4n3,Aguilar-Saavedra:2017nik}. The noncommutative space-time (NCST) is one of the features beyond the SM. We intend to investigate the polarized top quark decay in NCSM.

The non-commutativity in space-time is a possible generalization of the usual quantum mechanics and quantum field theory to describe the physics at very short distance of the order of the Plank length, since the nature of the space-time changes at these distances.
The properties of the top quark can be studied in noncommutative standard model (NCSM). Some before work are as following:

-At High energy collider experiments we can refer for example to  forbidden decays in standard model (SM) such as $Z\rightarrow\gamma\gamma$ \cite{Buric}, top quark decays \cite{ntwb,ntblv,n.mah}, compton scattering \cite{mathews}  which  have been investigated in NCST. In the experiment which has been done by  OPAL collaboration, NC bound from $e^- e^+$ scattering at 95\% CL is $\Lambda_{NC}> 141 GeV$ \cite{opal}. Also there are some researches in electron-proton colliders and hadron colliders to prob the NCST \cite{ep2017,Alboteanu:2006hh}

- Some studies is done on the bosons and fermions \cite{bosonnc1,bosonnc2,bosonnc4,bosonnc5,bosonnc6,bosonnc7}  and on the forbidden decays \cite{ncmfd,ncmzfd} in the NCST.

-The bounds on the NC scale ${\Lambda _{NC}}$ are estimated using the measurements of the unpolarized top quark two-body decay width, $t \to W{\kern 1pt} b$, and the W-boson polarization in top pair events from CDF experiment at tevatron. The bounds on ${\Lambda _{NC}}$ from the measured top quark width and W polarization are ${\Lambda _{NC}} \ge \,\;625\,GeV$ and ${\Lambda _{NC}} \ge \,\;1550\,GeV$, respectively  \cite{ntwb}.

- Decay $t \to W{\kern 1pt} b$ is studied in the NCSM and lowest contribution to the decay comes from the terms quadratic in the matrix describing the NC effects. The NC effects are found to be significant only for low values of the NC characteristics scale ( ${\Lambda _{NC}}\sim \,140\,GeV$) \cite{n.mah}.

-The lepton spectrum from the decay of a polarized top quark is studied in the  NCST, $t( \uparrow ) \to {\kern 1pt} {\kern 1pt} b{\ell ^ + }{\upsilon _\ell }$ , where the hadron vertex $t \to W{\kern 1pt} b$  is used up to the order ${\theta ^2}$. The ${\ell ^ + }$  spin correlation coefficient has a remarkable deviation from the SM for $\Lambda  \le 1\;TeV$. Also the total decay rate of polarized top quark is shown with respect to (w.r.t)  NC scale where the NC effects is negligible for ${\Lambda _{NC}} \ge \,\;625\,GeV$, but $\Gamma _A$, $\Gamma _B$ and $\Gamma _C$ are not calculated \cite{ntblv}. Also the NC corrected vertex is used just in the hadronic vertex in \cite{ntblv}. In this article we study the decay $t( \uparrow ) \to {\kern 1pt} {\kern 1pt} b{\ell ^ + }{\upsilon _\ell }$ in the NCST that SM  leptonic and hadronic vertices can be replaced by NC corrected vertices. Therefore we calculate $\Gamma _A$, $\Gamma _B$ and $\Gamma _C$ in NCST up to the order ${\theta ^2}$. Considering the top quark polarization,  $\Gamma _C$ could be measured in the future colliders. 

In the short and mid-term future, top quark studies will be mainly driven by the LHC experiments. However, detection of top quark properties will be an integral part of particle physics studies at any future facility. A possible future lepton collider such as ${e^ - }{e^ + }$ would be enable to do the unique top physics program far beyond the capabilities of the LHC. A ${e^ - }{e^ + }$ collider will allow us to study electroweak production of $t{\kern 1pt} \bar t$ pairs with no concurring QCD background. Therefore, precise measurements of top quark properties such as $\Gamma _C$ become possible. It should be noted that some studies have been done on polarized top quark to search in future ${e^ - }{e^ + }$  colliders \cite{GK,GKT,PSh}.

The organization of the paper is as follows: In sec. \ref{section.R2} we give a short introduction of NCSM . In sec. \ref{section.R3} the UPDR, PPDR and APDR for $t( \uparrow ) \to {\kern 1pt} {\kern 1pt} b{\ell ^ + }{\upsilon _\ell }$ are calculated in NCSM. The results of sec. \ref{section.R3} are expressed and discussed in sec. \ref{section.R4}. Finally, in sec. \ref{section.R5} we summarize our results and give our conclusion.

\section{Noncommutative space time (NCST)}\label{section.R2}

The shortcomings of existing theories in ordinary space-time in the expression of the some phenomena and even the a slight deviation of some others from precise laboratory results is leaded to search for new physics beyond the SM to justify these cases. One of the new theories, which began inclusive research on it since 1999, is theory of the field in NCST. It come back to the string theory where the end points of an open string show the noncommutative behavior. In NCST, the commutation relation between space-time coordinates versus ordinary space-time is a nonzero quantity \cite{nc1}:

\begin{equation}\label{eq1}
\left[{{{\hat x}^\mu },\,{\kern 1pt} {\kern 1pt} {{\hat x}^\nu }}\right] = i{\theta ^{\mu \nu }}
\end{equation}
�

where ${\theta ^{\mu \nu }}$  is called NC parameter and is a constant, real and antisymmetric tensor.

Eq.(\ref{eq1}) enters in string theory through the Weyl-Moyal star product \cite{nc1,Douglas:2001ba,nc2}:

\begin{equation}\label{eq2}
\begin{array}{l}
{\left( {f\; * g} \right)_{\left( x \right)}} = \exp \left( {\frac{i}{2}{\theta ^{\mu \nu }}\partial _\mu ^y{\kern 1pt} \partial _\nu ^z} \right)f\left( y \right)g\left( z \right){|_{y = z = x}}\\
 = f\left( x \right)g\left( x \right) + \frac{i}{2}{\kern 1pt} {\kern 1pt} {\theta ^{\mu \nu }}\left( {{\partial _\mu }{\kern 1pt} f\left( x \right)} \right)\left( {{\partial _\nu }{\kern 1pt} g\left( x \right)} \right) + O\left( {{\theta ^2}} \right)
\end{array}
\end{equation}

Writing the theory of the field in NCST from this way causes problems such as the charge quantization and etc \cite{nc3,nc4}. To solve these problems, two solutions are proposed. The first solution known as the Seiberg-Witten map \cite{nc5}, the NC fields are expanded in terms of ordinary fields \cite{nc6} and the symmetry group is similar the SM one. In the second solution, the symmetry group is larger than the SM one and using two higgs mechanisms, the symmetry of the group reduces to the SM symmetry group \cite{nc7}. In this paper, we use the Seiberg-Witten map. One of the important features of the theory of NC field is that in addition to the correction of the Feynman rules related to the existing vertices, it predicts new vertices that does not exist in the SM. In the following, we study the decay of the polarized top quark into a b-quark, anti-lepton and neutrino related to it, using the W-boson channel. To study this decay, we needed to corrected vertex in NCST for fermions with W-boson using Seiberg-Witten map, which is \cite{nc8}:

\begin{equation}\label{eq3}
\begin{array}{l}
\frac{{ie}}{{2\sqrt 2 \sin {\theta _w}}}\left( \begin{array}{l}
V_f^{(ij)}\\
V_f^{*(ij)}
\end{array} \right)\{ ({\gamma _\mu } - \frac{i}{2}{\theta _{\mu \nu \rho }}{k^\nu }{p^\mu })(1 - {\gamma _5})\\
 - \frac{i}{2}{\theta _{\mu \nu }}[\left( \begin{array}{l}
{m_{f_u^{(i)}}}\\
{m_{f_d^{(j)}}}
\end{array} \right)p_{in}^\nu (1 - {\gamma _5}) - \left( \begin{array}{l}
{m_{f_d^{(j)}}}\\
{m_{f_u^{(i)}}}
\end{array} \right)p_{out}^\nu (1 + {\gamma _5})]\}
\end{array}
\end{equation}
where $m$ is the fermion mass and ${\theta _{\mu \nu \rho }}$ is defined as follows:

\begin{equation}\label{eq4}
{\theta _{\mu \nu \rho }} = {\theta _{\mu \nu }}{\kern 1pt} {\gamma _\rho } + {\theta _{\rho \mu }}{\kern 1pt} {\gamma _\nu } + {\theta _{\nu \rho }}{\kern 1pt} {\gamma _\mu}
\end{equation}

In the Eq.(\ref{eq3}), $V_f^{\left( {ij} \right)}$ is CKM matrix element for quarks and is equal to 1 for leptons \cite{d.g,nc8}.

In this paper, by replacing the Eq.(\ref{eq3}) in the hadron and lepton vertices, we calculate the invariant amplitude of decay in NCST up to the order ${\theta ^2}$. It should be mentioned that the following identities are used in the calculations \cite{ntblv}:

\begin{equation}\label{eq5}
{A_\alpha }.{\theta ^{\alpha \beta }}.{B_\beta } = A\theta B = \vec \theta .\left( {\vec A \times \vec B} \right)
\end{equation}

\begin{equation}\label{eq6}
{A_\mu }{\theta ^{\mu \nu }}\theta _\nu ^\alpha {B_\alpha } = |\vec \theta {|^2}\left( {\vec A \times \vec B} \right) - \left( {\vec A\;.{\kern 1pt} \;\vec \theta } \right)\left( {\vec B\;.\;\vec \theta } \right)
\end{equation}

The NC field theories are not unitary for ${\theta ^{\mu 0}} \ne 0$ , therefore we choose ${\theta ^{\mu 0}} = 0$ in our calculations \cite{nc9}.

\section{Polarized top quark decay in noncommutative space-time}\label{section.R3}

The top quark has short lifetime  because of its large mass. This quark decays before hadronization and its spin information can be deduced from its decay products. If polarized top quark decays in the form of semileptonic so that b quark and ${W^ + }$- boson to be decay products and ${W^ + }$ decays into a charged lepton ${\ell ^ + }$ and a neutrino ${\upsilon _\ell }$, so we have $\,t({p_1}) \to b({p_2})\,{W^ + }(q) \to b({p_2})\,\,{\ell ^ + }({p_3})\,\,{\upsilon _\ell }({p_4})\,$. Here $t$, $b$, ${\ell ^ + }$ and ${\upsilon _\ell }$ particles have been tagged with 1, 2, 3 and 4 numbers, respectively.  ${W^ + }$- boson four-vector is $q = {p_3} + {p_4}$.

\begin{figure}[t]
  \centerline{\includegraphics[width=0.45\textwidth]{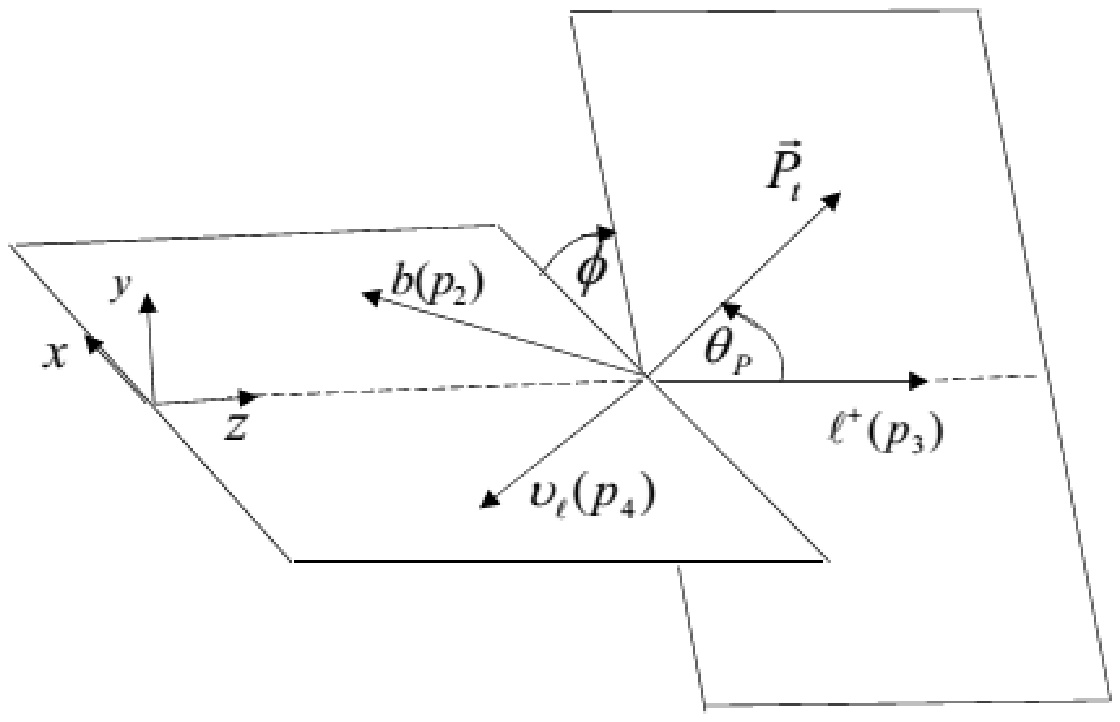}}
  \caption{The definition of the polar angle ${\theta _P}$ and the azimuthal angle $\phi $ in the rest frame decay of a polarized top quark. The event plane defines the (x,z) plane with ${{\vec p}_\ell }||{\kern 1pt} \,z$.}
  \label{fig.0}
\end{figure}

Considering helicity system shown in Fig \ref{fig.0}  that the event plane defines the (x,z) plane and the z-axis be in the direction of the ${\ell ^ + }$ momentum and also considering top quark polarization vector, ${\vec P_t}$, so that makes ${\theta _P}$ polar angle with ${\ell ^ + }$ momentum and $\phi $ azimuthal angle with the event plane(Figure 1 in Ref.\cite{base}), we can write $\,t({p_1}) \to b({p_2})\,{W^ + }(q) \to b({p_2})\,\,{\ell ^ + }({p_3})\,\,{\upsilon _\ell }({p_4})\,$ decay amplitude for LO in NCSM by using the Seiberge-Witten map as follows:
\begin{equation}\label{eq7}
\begin{array}{l}
{M^{NC}} = \left( {{{(\frac{{ - i{g_w}}}{{\sqrt 2 }})}^2}{V_{tb}}} \right)\\
\\
 \times \,\bar u\left( {{p_2},{s_2}} \right)\left\{ {{\gamma _\mu }\left( {\frac{{1 - {\gamma _5}}}{2}} \right) + \frac{i}{2}{\theta _{\mu \alpha \rho }}{q^\alpha }p_1^\rho \left( {\frac{{1 - {\gamma _5}}}{2}} \right)} \right\}u\left( {{p_1},{s_1}} \right)\\
\\
 \times \left( { - i\frac{{{g^{\mu \nu }} - \frac{{{q^\mu }{q^\nu }}}{{m_W^2}}}}{{{q^2} - m_W^2 + i\varepsilon }}} \right)\\
\\
 \times \,\bar u\left( {{p_4},{s_4}} \right)\left\{ {{\gamma _\nu }\left( {\frac{{1 - {\gamma _5}}}{2}} \right) - \frac{i}{2}{\theta _{\nu \beta \sigma }}{q^\beta }p_3^\sigma \left( {\frac{{1 - {\gamma _5}}}{2}} \right)} \right\}\upsilon \left( {{p_3},{s_3}} \right)
\end{array}
\end{equation}

LO in NCST means that we calculate the amplitude in the LO of SM while vertices are replaced by the corrected one in the NCST. To obtain the amplitude(Eq.(\ref{eq7})), we have used the SM ${W^ + }$ propagator (because propagators don't receive any correction in NCST) and Eq.(\ref{eq3}) for the corrected fermion-fermion-W boson in NCST. Also lepton masses and b quark mass are neglected.

We write W-propagator as \cite{diss}
\begin{equation}\label{eq07}
\begin{array}{l}
 - i\frac{{{g_{\mu \nu }}}}{{{q^2} - m_W^2 + i\varepsilon }} \to  - i\frac{{{g_{\mu \nu }}}}{{{q^2} - m_W^2 + i{m_W}{\Gamma _W}}}\\
\\
 \to  - i\frac{{{g_{\mu \nu }}}}{{m_W^2}}\frac{1}{{{{{y^2}} \mathord{\left/
 {\vphantom {{{y^2}} {{{\hat y}^2}}}} \right.
 \kern-\nulldelimiterspace} {{{\hat y}^2}}} - 1 + i\gamma }}
\end{array}
\end{equation}

 Eq.(\ref{eq7}) could be written  as follows:
\begin{equation}\label{eq8}
{M^{NC}} = i\frac{{{G_F}{\kern 1pt} {V_{tb}}}}{{\sqrt 2 {\kern 1pt} (\frac{{{y^2}}}{{{{\hat y}^2}}} - 1 + i\gamma )}}H_\mu ^{NC}L_{NC}^\mu
\end{equation}

where $y = \sqrt {\frac{{{q^2}}}{{m_1^2}}}$, $\hat y = \frac{{{m_W}}}{{{m_1}}}$, $\gamma  = \frac{{{\Gamma _W}}}{{{m_W}}}$ and $q$ and ${\Gamma _W}$ are ${W}$- boson momentum and resonance width. The hadronic current $H_\mu ^{NC}$ and leptonic current $L_\mu ^{NC}$ in NCST and ${G_F}$ Fermi coupling constant are defined as follows:

\begin{equation}\label{eq9}
\begin{array}{l}
H_\mu ^{NC} = \bar u\left( {{p_2},{s_2}} \right)\{ {\gamma _\mu }\left( {1 - {\gamma _5}} \right)\\
\\
 + \frac{i}{2}{\theta _{\mu {\kern 1pt} \alpha \rho }}{\kern 1pt} {q^\alpha }p_1^\rho \left( {1 - {\gamma _5}} \right)\} u\left( {{p_1},{s_1}} \right)
\end{array}
\end{equation}

\begin{equation}\label{eq10}
\begin{array}{l}
L_\mu ^{NC} = \bar u\left( {{p_4},{s_4}} \right)\{ {\gamma _\mu }\left( {1 - {\gamma _5}} \right)\\
\\
 - \frac{i}{2}{\theta _{\mu \beta \sigma }}{\kern 1pt} {q^\beta }p_3^\sigma \left( {1 - {\gamma _5}} \right)\} \upsilon \left( {{p_3},{s_3}} \right)
\end{array}
\end{equation}

\begin{equation}\label{eq11}
{G_F} = \frac{{g_w^2}}{{4\sqrt 2 {\kern 1pt} m_W^2}}
\end{equation}

We need the average of  squared amplitude to obtain the decay rate. Therefore with averaging over initial spins, sum over final spins and using Casimir's trick we calculate it( $\left\langle {{{\left| {{M^{NC}}} \right|}^2}} \right\rangle$). Also we have defined NC vector, ${\vec \theta }$, in the (x,z) plane (${\theta ^{31}} = 0$) and $\vec \theta  = \frac{{|\vec \theta |}}{{\sqrt 2 }}\left( {1{\kern 1pt} {\kern 1pt} ,0{\kern 1pt} {\kern 1pt} ,1} \right)$ where $|\vec \theta | = \theta  \sim \frac{1}{{\Lambda _{NC}^2}}$ (${\Lambda _{NC}}$ is the NC scale). We have

 �
\begin{equation}\label{eq12}
\begin{array}{l}
\left\langle {{{\left| {{M^{NC}}} \right|}^2}} \right\rangle  = \frac{{G_F^{{\kern 1pt} 2}{{\left| {\;{V_{tb}}} \right|}^2}}}{{2\,[\;{{\left( {1 - \frac{{{y^2}}}{{{{\hat y}^2}}}} \right)}^2} + {\gamma ^2}]}}\\
 \times \;\{ \;128\left( {{{\tilde p}_1}.{p_3}} \right)\left( {{p_2}.{p_4}} \right)\\
 - \;16\left( {q.\theta .\theta .q} \right)\left( {{{\tilde p}_1}.{p_1}} \right)\left( {{p_1}.{p_2}} \right)\left( {{p_3}.{p_4}} \right)\\
 + \;8\left( {q.\theta .\theta .q} \right)\left( {{{\tilde p}_1}.{p_2}} \right)p_1^2\left( {{p_3}.{p_4}} \right)\\
 + \;32\left( {{{\tilde p}_1}.{p_1}} \right)p_3^0{q^1}{\theta ^{21}}p_4^1p_2^3\; - \,32\left( {{{\tilde p}_1}.{p_1}} \right)p_3^0{q^1}{\theta ^{21}}p_4^3p_2^1\\
 + 32\left( {{{\tilde p}_1}.{p_1}} \right)p_3^0{q^3}{\theta ^{23}}p_4^1p_2^3 - \;32\left( {{{\tilde p}_1}.{p_1}} \right)p_3^0{q^3}{\theta ^{23}}p_4^3p_2^1\\
 + 32\left( {{{\tilde p}_1}.{p_1}} \right)p_4^0{q^1}{\theta ^{21}}p_3^3p_2^1 + 32\left( {{{\tilde p}_1}.{p_1}} \right)p_4^0{q^3}{\theta ^{23}}p_3^3p_2^1\\
 - \;32\left( {{{\tilde p}_1}.{p_1}} \right)p_2^0{q^1}{\theta ^{21}}p_3^3p_4^1 - \;32\left( {{{\tilde p}_1}.{p_1}} \right)p_2^0{q^3}{\theta ^{23}}p_3^3p_4^1\\
 + \;32\left( {{{\tilde p}_1}.{p_2}} \right)p_1^0{q^1}{\theta ^{21}}p_3^3p_4^1 + 32\left( {{{\tilde p}_1}.{p_2}} \right)p_1^0{q^3}{\theta ^{23}}p_3^3p_4^1\\
 + \;32\left( {{p_1}.{p_2}} \right)p_3^0{q^1}{\theta ^{21}}p_4^1\tilde p_1^3 - \;32\left( {{p_1}.{p_2}} \right)p_3^0{q^1}{\theta ^{21}}p_4^3\tilde p_1^1\\
 + \;32\left( {{p_1}.{p_2}} \right)p_3^0{q^3}{\theta ^{23}}p_4^1\tilde p_1^3 - \;32\left( {{p_1}.{p_2}} \right)p_3^0{q^3}{\theta ^{23}}p_4^3\tilde p_1^1\\
 + \;32\left( {{p_1}.{p_2}} \right)p_4^0{q^1}{\theta ^{21}}p_3^3\tilde p_1^1 + \;32\left( {{p_1}.{p_2}} \right)p_4^0{q^3}{\theta ^{23}}p_3^3\tilde p_1^1\\
 - \;32\left( {{p_1}.{p_2}} \right)\tilde p_1^0{q^1}{\theta ^{21}}p_3^3p_4^1 - \;32\left( {{p_1}.{p_2}} \right)\tilde p_1^0{q^3}{\theta ^{23}}p_3^3p_4^1\\
 + \;32\left( {{p_3}.{p_4}} \right)p_1^0{q^1}\tilde p_1^1{\theta ^{21}}p_2^3 - 32\left( {{p_3}.{p_4}} \right)p_1^0{q^1}\tilde p_1^3{\theta ^{21}}p_2^1\\
 + 32\left( {{p_3}.{p_4}} \right)p_1^0{q^3}\tilde p_1^1{\theta ^{23}}p_2^3 - 32\left( {{p_3}.{p_4}} \right)p_1^0{q^3}\tilde p_1^3{\theta ^{23}}p_2^1\}
\end{array}
\end{equation}

In Eq.(\ref{eq12}), the momentum four-vectors and the top quark polarization vector($s^\mu$) in the rest frame of top quark are

\[{p_1^\mu} ={p_t^\mu}= {m_1}(1;\,0,\,0,\,0),\]
\[{p_2^\mu} = {p_b^\mu} = \frac{{{m_1}}}{2}(1 - {y^2})(1;\,\sin {\theta _b},\,0,\,\cos {\theta _b}),\]
\[{p_3^\mu} = {p_{\ell^+}^\mu } = \frac{{{m_1}}}{2}x(1;\,0,\,0,\,1),\]
\[{p_4^\mu} = {p_\nu^\mu} = \frac{{{m_1}}}{2}(1 - x + {y^2})(1;\, - \sin {\theta _\upsilon },\,0,\,\cos {\theta _\upsilon }),\]
\[{s_1^\mu}={s_t^\mu} = P(0;\,\sin {\theta _P}\,\cos \phi ,\,\sin {\theta _P}\,\sin \phi ,\,\cos {\theta _P}\,),\]
\[{\tilde p_1^\mu} ={p_1^\mu}-{m_1 s_1^\mu}\]

where
\[\cos {\theta _\upsilon } = \frac{{x(1 - x + {y^2}) - 2{y^2}}}{{x(1 - x + {y^2})}},\]
\[\cos {\theta _b} = \frac{{2{y^2} - x(1 + {y^2})}}{{x(1 - {y^2})}}.\]

Using 4-vectors and Eqs.(\ref{eq5})-(\ref{eq6}) we can calculate Eq.(\ref{eq12}).

For the contraction of the lepton and hadron tensors in NCST, we have
\begin{equation}\label{eq13}
\begin{array}{l}
H_{\mu {\kern 1pt} \nu }^{(NC)}L_{(NC)}^{\mu \nu } = 32m_1^4[M_{(NC)}^A(x,y)\\
\\
 + M_{(NC)}^B(x,y)\cos {\theta _P} + M_{(NC)}^C(x,y)\sin {\theta _P}\cos \phi ]
\end{array}
\end{equation}

Where $M_{(NC)}^A$, $M_{(NC)}^B$ and $M_{(NC)}^C$ are the unpolarized, polar polarized and azimuthal polarized amplitudes, respectively and we calculate them in the following. It should be noted that the scaled lepton energy $x$ is equal to $x = \frac{{2{E_\ell }}}{{{m_t}}}$.

For the unpolarized amplitude, we have
\begin{equation}\label{eq14}
\begin{array}{l}
M_{(NC)}^A\left( {x,y} \right) = \;x\left( {1 - x} \right)\\
 + \frac{1}{{8\sqrt 2 }}|\vec \theta |m_1^2\left( {x\left( {1 + {y^2}} \right) - 2{y^2}} \right)\left( {2 \times \frac{{\sqrt {{y^2}\left( {x - {x^2} - {y^2} + x{y^2}} \right)} }}{x}} \right)\\
 + \frac{1}{{8\sqrt 2 }}|\vec \theta |m_1^2\left( {\frac{{4{y^2}\left( {x - {x^2} - {y^2} + x{y^2}} \right)}}{x}} \right)\;\;\\
 - \;\frac{1}{{128}}|\vec \theta {|^2}m_1^4\;{y^2}{\left( {1 - {y^2}} \right)^3}\\
 + \frac{1}{{128}}|\vec \theta {|^2}m_1^4{y^2}\left( {1 - {y^2}} \right)\left( {\frac{{2{y^2} - x\left( {1 + {y^2}} \right)}}{x}} \right)\\
 \times \left( {4 \times \frac{{\;\sqrt {{y^2}\left( {x - {x^2} - {y^2} + x{y^2}} \right)} }}{x}} \right)\;,
\end{array}
\end{equation}

First term in Eq.(\ref{eq14}) shows the ordinary space-time contribution and other terms of order $\theta $ and ${\theta ^2}$ are the corrections caused by noncommutativity.

The polar polarized amplitude is
\begin{equation}\label{eq15}
\begin{array}{l}
M_{(NC)}^B\left( {x,y} \right) = \;x\left( {1 - x} \right)\\
 + \frac{1}{{8\sqrt 2 }}|\vec \theta |m_1^2\left( {\frac{{4{y^2}\left( {x - {x^2} - {y^2} + x{y^2}} \right)}}{x}} \right)\\
 - \frac{1}{{8\sqrt 2 }}|\vec \theta |m_1^2\left( {2{y^2} - x\left( {1 + {y^2}} \right)} \right)\left( {2 \times \frac{{\,\sqrt {{y^2}\left( {x - {x^2} - {y^2} + x{y^2}} \right)} }}{x}} \right)\\
 + \frac{1}{{128}}|\vec \theta {|^2}m_1^4{y^2}{\left( {1 - {y^2}} \right)^2}\left( {\frac{{2{y^2} - x\left( {1 + {y^2}} \right)}}{x}} \right)\\
 - \frac{1}{{128}}|\vec \theta {|^2}m_1^4{y^2}{\left( {\frac{{2{y^2} - x\left( {1 + {y^2}} \right)}}{x}} \right)^2}\left( {4 \times \frac{{\;\sqrt {{y^2}\left( {x - {x^2} - {y^2} + x{y^2}} \right)} }}{x}} \right)
\end{array}
\end{equation}

Eq.(\ref{eq15}) expresses the contributions of the corrections caused by noncommutativity in order $\theta $ and ${\theta ^2}$ that they are added to the ordinary space-time contribution when $\theta  = 0$.
Eventually for the azimuthal polarized amplitude, we have
\begin{equation}\label{eq16}
\begin{array}{l}
M_{(NC)}^C\left( {x,y} \right) = \\
 - \;\frac{1}{{4\sqrt 2 }}|\vec \theta |m_1^2{y^2}\left( {1 - x} \right)\left( {2 \times \frac{{\sqrt {{y^2}\left( {x - {x^2} - {y^2} + x{y^2}} \right)} }}{x}} \right)\\
 - \;\frac{1}{{4\sqrt 2 }}|\vec \theta |m_1^2{y^2}\left( {1 - x} \right)\left( {\frac{{x\left( {1 + {y^2}} \right) - 2{y^2}}}{x}} \right)\\
\frac{1}{{128}}|\vec \theta {|^2}m_1^4{y^2}{\left( {1 - {y^2}} \right)^2}\left( {2 \times \frac{{\sqrt {{y^2}\left( {x - {x^2} - {y^2} + x{y^2}} \right)} }}{x}} \right)\\
 - \;\frac{1}{{128}}|\vec \theta {|^2}m_1^4{y^2}\left( {\frac{{2{y^2} - x\left( {1 + {y^2}} \right)}}{x}} \right)\left( {\frac{{8{y^2}\left( {x - {x^2} - {y^2} + x{y^2}} \right)}}{{{x^2}}}} \right)
\end{array}
\end{equation}

As mentioned in the introduction the azimuthal polarized amplitude is zero in the LO of SM ($M_0^C = 0$ see Refs.\cite{base,diss}), therefore Eq.(\ref{eq16}) shows the NC contribution of order $\theta $ and ${\theta ^2}$.
The differential rate is \cite{d.g}
\begin{equation}\label{eq17}
\begin{array}{l}
d{\Gamma ^{NC}} = \frac{1}{{2{m_1}}}\frac{1}{{{{(2\pi )}^3}}}\frac{{{d^3}{{\vec p}_2}}}{{2{E_2}}}\frac{1}{{{{(2\pi )}^3}}}\frac{{{d^3}{{\vec p}_3}}}{{2{E_3}}}\frac{1}{{{{(2\pi )}^3}}}\frac{{{d^3}{{\vec p}_4}}}{{2{E_4}}}\\
\\
\quad \quad \;\; \times {(2\pi )^4}{\delta ^4}({p_1} - {p_2} - {p_3} - {p_4})\overline {{{\left| {{M^{NC}}} \right|}^2}}
\end{array}
\end{equation}

Subestituting the Eq.(\ref{eq13}) in the Eq.(\ref{eq17}), the differential decay rate is given by
\begin{equation}\label{eq18}
\begin{array}{l}
\frac{{d{\Gamma ^{NC}}}}{{dxd\cos {\theta _P}d\phi }} = \frac{1}{{4\pi }}\{ \,{\Gamma _F}\frac{{12}}{{{{(1 - \frac{{{y^2}}}{{{{\hat y}^2}}})}^2} + {\gamma ^2}}}d{y^2}[M_{(NC)}^A(x)\\
\\
 + M_{(NC)}^B(x)\cos {\theta _P} + M_{(NC)}^C(x)\sin {\theta _P}\cos \phi ]\}
\end{array}
\end{equation}
The general form of the differential decay rate is:
\begin{equation}\label{eq19}
\begin{array}{l}
\frac{{d{\Gamma ^{NC}}}}{{dxd\cos {\theta _P}d\phi }} = \frac{1}{{4\pi }}[\frac{{d\Gamma _A^{NC}}}{{dx}} + \frac{{d\Gamma _B^{NC}}}{{dx}}\cos {\theta _P}\\
\\
 + \frac{{d\Gamma _C^{NC}}}{{dx}}\sin {\theta _P}\cos \phi ]
\end{array}
\end{equation}

By inserting Eqs.(\ref{eq18}) and (\ref{eq19}) and using the narrow-width approximation for the W-propagator (see sec.\ref{section.R1}), we obtain the differential rates in NCST i.e. the unpolarized ($\frac{{d\Gamma _A^{NC}}}{{dx}}$), the polar polarized ($\frac{{d\Gamma _B^{NC}}}{{dx}}$) and the azimuthal polarized ($\frac{{d\Gamma _C^{NC}}}{{dx}}$) differential rates. The important point is that the differential rates dependence on y disappears. They are only depend on x, as in ordinary space-time.
Therefore the unpolarized differential rate (UPDiR) is
\begin{equation}\label{eq20}
\frac{{d\Gamma _A^{NC}}}{{dx}} = \frac{{G_F^2m_1^3m_W^3}}{{16{\pi ^2}{\Gamma _W}}}{\left| {{\kern 1pt} {V_{tb}}} \right|^2}\{ {F_0}(x) + \frac{1}{{{\Lambda ^2}}}{F_1}(x) + \frac{1}{{{\Lambda ^4}}}{F_2}(x)\}
\end{equation}
Where
\begin{equation}\label{eq21}
\begin{array}{l}
{F_0}(x) = x(1 - x),\\
\\
 {F_1}(x) = \frac{1}{{8\sqrt 2 }}m_1^2(4{{\hat y}^2}\frac{{ - {x^2} + (1 + {{\hat y}^2})x - {{\hat y}^2}}}{x}\\
\qquad\qquad + 2\hat y(x(1 + {{\hat y}^2}) - 2{{\hat y}^2})\frac{{\sqrt { - {x^2} + (1 + {{\hat y}^2})x - {{\hat y}^2}} }}{x}),\\
\\
 {F_2}(x) = \frac{1}{{128}}m_1^4( - {{\hat y}^2}{(1 - {{\hat y}^2})^3}\\
 \qquad\qquad + 4{{\hat y}^3}(1 - {{\hat y}^2})(2{{\hat y}^2} - x(1 + {{\hat y}^2}))\frac{{\sqrt { - {x^2} + (1 + {{\hat y}^2})x - {{\hat y}^2}} }}{{{x^2}}}).
\end{array}
\end{equation}

First term in Eq.(\ref{eq20}), ${F_0}(x)$, is ordinary space-time contribution and the other terms of order $\theta  \sim \frac{1}{{{\Lambda ^2}}}$ and ${\theta ^2} \sim \frac{1}{{{\Lambda ^4}}}$, ${F_1}(x)$ and ${F_2}(x)$, are NCST contributions.

The polar polarized differential rate (PPDiR) is
\begin{equation}\label{eq22}
\frac{{d\Gamma _B^{NC}}}{{dx}} = \frac{{G_F^2m_1^3m_W^3}}{{16{\pi ^2}{\Gamma _W}}}{\left| {{\kern 1pt} {V_{tb}}} \right|^2}\{ {G_0}(x) + \frac{1}{{{\Lambda ^2}}}{G_1}(x) + \frac{1}{{{\Lambda ^4}}}{G_2}(x)\}
\end{equation}

Where
\begin{equation}\label{eq23}
\begin{array}{l}
{G_0}(x) = x(1 - x),\\
\\
 {G_1}(x) = \frac{1}{{8\sqrt 2 }}m_1^2(4{{\hat y}^2}\frac{{ - {x^2} + (1 + {{\hat y}^2})x - {{\hat y}^2}}}{x}\\
 \qquad\qquad + 2\hat y(x(1 + {{\hat y}^2}) - 2{{\hat y}^2})\frac{{\sqrt { - {x^2} + (1 + {{\hat y}^2})x - {{\hat y}^2}} }}{x}),\\
\\
 {G_2}(x) = \frac{1}{{128}}m_1^4({{\hat y}^2}{(1 - {{\hat y}^2})^2}(\frac{{2{{\hat y}^2} - x(1 + {{\hat y}^2})}}{x})\\
\qquad\qquad  - 4{{\hat y}^3}{(2{{\hat y}^2} - x(1 + {{\hat y}^2}))^2}\frac{{\sqrt { - {x^2} + (1 + {{\hat y}^2})x - {{\hat y}^2}} }}{{{x^3}}}).
\end{array}
\end{equation}

Like the unpolarized case, Eq.(\ref{eq22}) contains the ordinary space-time  contribution in order of $\theta  = 0$ and NCST ones in order of $\theta  \sim \frac{1}{{{\Lambda ^2}}}$ and ${\theta ^2} \sim \frac{1}{{{\Lambda ^4}}}$.

The azimuthal polarized differential rate (APDiR) in NCST is

\begin{equation}\label{eq24}
\frac{{d\Gamma _C^{NC}}}{{dx}} = \frac{{G_F^2m_1^3m_W^3}}{{16{\pi ^2}{\Gamma _W}}}{\left| {{\kern 1pt} {V_{tb}}} \right|^2}\{ \frac{1}{{{\Lambda ^2}}}{H_1}(x) + \frac{1}{{{\Lambda ^4}}}{H_2}(x)\}
\end{equation}

Where
\begin{equation}\label{eq25}
\begin{array}{l}
{H_1}(x) = \frac{1}{{4\sqrt 2 }}m_1^2( - 2{{\hat y}^3}(1 - x)\frac{{\sqrt { - {x^2} + (1 + {{\hat y}^2})x - {{\hat y}^2}} }}{x},\\
 \qquad  \qquad- {{\hat y}^2}(1 - x)\frac{{(1 + {{\hat y}^2})x - 2{{\hat y}^2}}}{x}),\\
\\
 {H_2}(x) = \frac{1}{{128}}m_1^4[2{{\hat y}^3}{(1 - {{\hat y}^2})^2}{\kern 1pt} {\kern 1pt}  \times \frac{{\sqrt { - {x^2} + (1 + {{\hat y}^2})x - {{\hat y}^2}} }}{x}\\
  \qquad\qquad - 8{{\hat y}^4}(\frac{{ - (1 + {{\hat y}^2})x + 2{{\hat y}^2}}}{x})(\frac{{ - {x^2} + (1 + {{\hat y}^2})x - {{\hat y}^2}}}{{{x^2}}})].
\end{array}
\end{equation}

As mentioned before, the LO SM contribution for APDiR is zero and Eq.(\ref{eq24}) contains the NCST contributions only. The integrated rates are calculated in the  interval ${\hat y^2} \le x \le 1$. We have:
\begin{equation}\label{eq26}
\begin{array}{l}
\Gamma _A^{NC} = \frac{{G_F^2m_1^3m_W^3}}{{16{\pi ^2}{\Gamma _W}}}{\left| {{\kern 1pt} {V_{tb}}} \right|^2}\{ \frac{1}{6}{(1 - {{\hat y}^2})^2}(1 + 2{{\hat y}^2})\\
\\
 + \frac{1}{{{\Lambda ^2}}}[\frac{1}{{8\sqrt 2 }}m_1^2(2{{\hat y}^2}(1 - {{\hat y}^4}) + 8{{\hat y}^4}\ln \hat y\\
\\
 - \frac{1}{4}\hat y(1 + {{\hat y}^2}){(1 - {{\hat y}^2})^2}\pi  + 2{{\hat y}^3}{(1 - \hat y)^2}\pi )]\\
\\
 + \frac{1}{{{\Lambda ^4}}}[\frac{1}{{128}}m_1^4( - {{\hat y}^2}{(1 - {{\hat y}^2})^4} + 2{{\hat y}^3}(1 - {{\hat y}^2}){(1 - \hat y)^4}\pi )]\}
\end{array}
\end{equation}
 for the unpolarized rate,
\begin{equation}\label{eq27}
\begin{array}{l}
\Gamma _B^{NC} = \frac{{G_F^2m_1^3m_W^3}}{{16{\pi ^2}{\Gamma _W}}}{\left| {{\kern 1pt} {V_{tb}}} \right|^2}\{ \frac{1}{6}{(1 - {{\hat y}^2})^2}(1 + 2{{\hat y}^2})\\
\\
 + \frac{1}{{{\Lambda ^2}}}[\frac{1}{{8\sqrt 2 }}m_1^2(2{{\hat y}^2}(1 - {{\hat y}^4}) + 8{{\hat y}^4}\ln \hat y\\
\\
 - \frac{1}{4}\hat y(1 + {{\hat y}^2}){(1 - {{\hat y}^2})^2}\pi  + 2{{\hat y}^3}{(1 - \hat y)^2}\pi )]\\
\\
 + \frac{1}{{{\Lambda ^4}}}[\frac{1}{{128}}m_1^4( - 4{{\hat y}^4}{(1 - {{\hat y}^2})^2}\ln \hat y\\
\\
 - {{\hat y}^2}(1 + {{\hat y}^2}){(1 - {{\hat y}^2})^3} - 8{{\hat y}^6}\pi  + 16{{\hat y}^5}(1 + {{\hat y}^2})\pi \\
\\
 - 10{{\hat y}^4}{(1 + {{\hat y}^2})^2}\pi  + 2{{\hat y}^3}{(1 + {{\hat y}^2})^3}\pi )]\}
\end{array}
\end{equation}
for the polar polarized rate and
\begin{equation}\label{eq28}
\begin{array}{l}
\Gamma _C^{NC} = \frac{{G_F^2m_1^3m_W^3}}{{16{\pi ^2}{\Gamma _W}}}{\left| {{\kern 1pt} {V_{t{\kern 1pt} b}}} \right|^2}\\
\\
\{ \frac{1}{{{\Lambda ^2}}}[\frac{1}{{4\sqrt 2 }}m_1^2( - \frac{1}{4}{{\hat y}^3}{(1 - {{\hat y}^2})^2}\pi  + {{\hat y}^3}{(1 - \hat y)^2}\pi \\
\quad \quad \quad \quad \quad \quad \quad \quad \quad \\
 - \frac{1}{2}{{\hat y}^2}({(1 - \hat y)^4} + 4{{\hat y}^2})(1 - {{\hat y}^2}) - 4{{\hat y}^2}\ln \hat y)]\\
\quad \quad \quad \quad \quad \quad \quad \quad \quad \\
 + \frac{1}{{{\Lambda ^4}}}[\frac{1}{{128}}m_1^4( - {{\hat y}^3}{(1 - {{\hat y}^2})^2}{(1 - \hat y)^2}\pi  - 24{{\hat y}^4}(1 - {{\hat y}^4})\\
\quad \quad \quad \quad \quad \quad \quad \quad \quad \\
 - 16{{\hat y}^4}(1 + 4{{\hat y}^2} + {{\hat y}^4})\ln \hat y)]\}
\end{array}
\end{equation}
for the azimuthal polarized rate.

One can takes the variables as ${m_1} = 173.21{\kern 1pt} {\kern 1pt} \,\,GeV$, ${m_W} = 80.385{\kern 1pt} {\kern 1pt} \,\;GeV$, $\hat y = \frac{{{m_W}}}{{{m_1}}}$, ${\Gamma _W} = 2.085\,\,\,GeV$, ${G_F} = 1.1663787 \times {10^{ - 5}}{\kern 1pt} \,{(GeV)^{ - 2}}$ and ${V_{tb}} = 0.999$ \cite{pdg},  $\Lambda $. We obtain Eqs.(\ref{eq26} -\ref{eq28}) w.r.t NC scale, by replacing the mentioned variables, as follows:

The unpolarized rate :
\begin{equation}\label{eq29}
\Gamma _A^{NC} = \Gamma _A^{(0)} + \Gamma _A^{n{\kern 1pt} c} = 1.11308(0.146803 + \frac{{88.9289}}{{{\Lambda ^2}}}- \frac{{288195}}{{{\Lambda ^4}}})
\end{equation}
$\Gamma _A^{{\kern 1pt} {\kern 1pt} (0)}$ is the unpolarized rate in LO of SM and $\Gamma _A^{{\kern 1pt} {\kern 1pt} (nc)}$ is the contribution of NCST in order of $\frac{1}{{{\Lambda ^2}}}$ and $\frac{1}{{{\Lambda ^4}}}$.

The polar polarized rate:
\begin{equation}\label{eq30}
\Gamma _B^{NC} = \Gamma _B^{(0)} + \Gamma _B^{n{\kern 1pt}c} = 1.11308(0.146803 + \frac{{88.9289}}{{{\Lambda ^2}}}+ \frac{{1189.41}}{{{\Lambda ^4}}} )
\end{equation}
Where $\Gamma _B^{{\kern 1pt} {\kern 1pt} (0)}$ is the polar polarized rate in LO of SM and $\Gamma _B^{{\kern 1pt} {\kern 1pt} (nc)}$ is the contribution of NCST in order of $\frac{1}{{{\Lambda ^2}}}$ and $\frac{1}{{{\Lambda ^4}}}$.

The azimuthal polarized rate:
\begin{equation}\label{eq31}
\Gamma _C^{NC} = 1.11308(\frac{{164.045}}{{{\Lambda ^2}}} - \frac{{211731}}{{{\Lambda ^4}}})
\end{equation}

The contribution of azimuthal polarized rate in LO of SM is zero and Eq.(\ref{eq31}) shows the azimuthal polarized rate in NCST of order $\frac{1}{{{\Lambda ^2}}}$ and $\frac{1}{{{\Lambda ^4}}}$. It has the non-zero value.

According to theoretical calculations of rates, if an experiment's central values exactly agree with the SM prediction for $\Gamma _A$, $\Gamma _B$ and $\Gamma _C$ and etc, and also the anticipated errors are small and ignorable, then obtained limits in next section are acceptable. Otherwise, these limits depend on the amount of experimental errors.

\section{Results and Discussions}\label{section.R4}

\begin{figure}[t]
  \centerline{\includegraphics[width=0.5\textwidth]{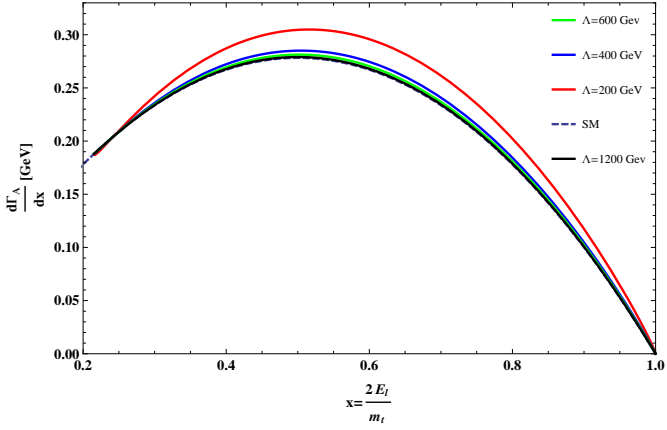}}
  \caption{The unpolarized differential rates w.r.t the charged lepton energy(charged lepton spectra).}
  \label{fig.1} \end{figure}

The NC effects on the decay of polarized top quark has been investigated in sec.\ref{section.R3} by inserting the Eq.(\ref{eq3}) in the hadronic and leptonic vertices. The corrections of the NCST has been calculated up to LO in terms of the NC parameter, $\theta $. The UPDiR has been calculated in NCST up to the order of ${\theta ^2} \sim \frac{1}{{{\Lambda ^4}}}$  in Eq.(\ref{eq20}). The UPDiR are depicted in Fig.\ref{fig.1}. The value of UPDiR is changing by variation of NC scale. For $\Lambda  \ge 1200\;GeV$ the variation is not recognizable with its SM value.

\begin{figure}[t]
 \centerline{\includegraphics[width=0.5\textwidth]{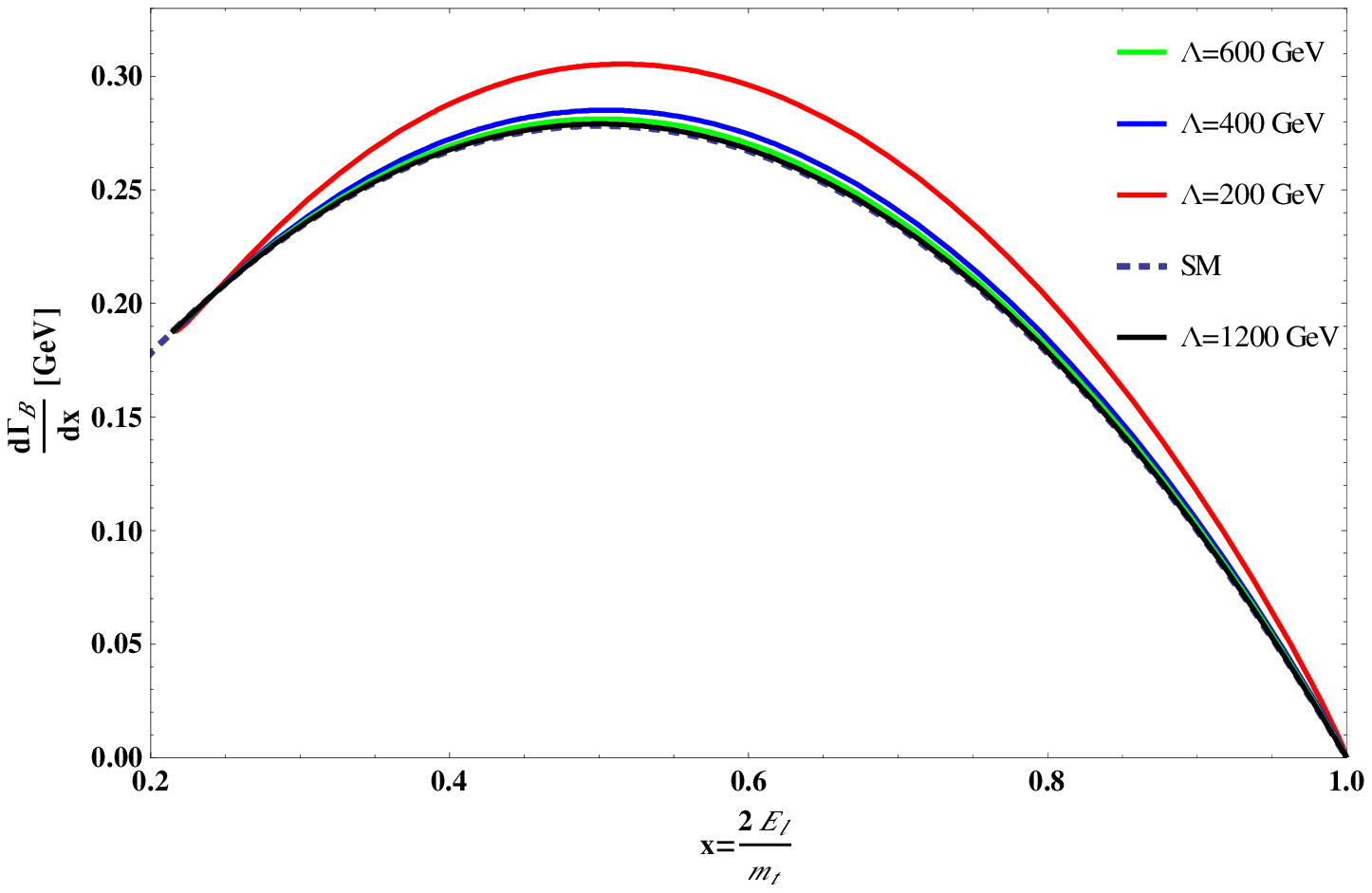}}   \caption{The polar polarized differential rates w.r.t the charged lepton energy(charged lepton spectra).}  \label{fig.2} \end{figure}

The PPDiR has been calculated in NCST up to the order of ${\theta ^2} \sim \frac{1}{{{\Lambda ^4}}}$ in Eq.(\ref{eq22}). In Fig \ref{fig.2} the PPDiR is depicted for different values of $\Lambda $. The variation from the SM value decreases when the $\Lambda $ is increasing.

\begin{figure}[t]
 \centerline{\includegraphics[width=0.5\textwidth]{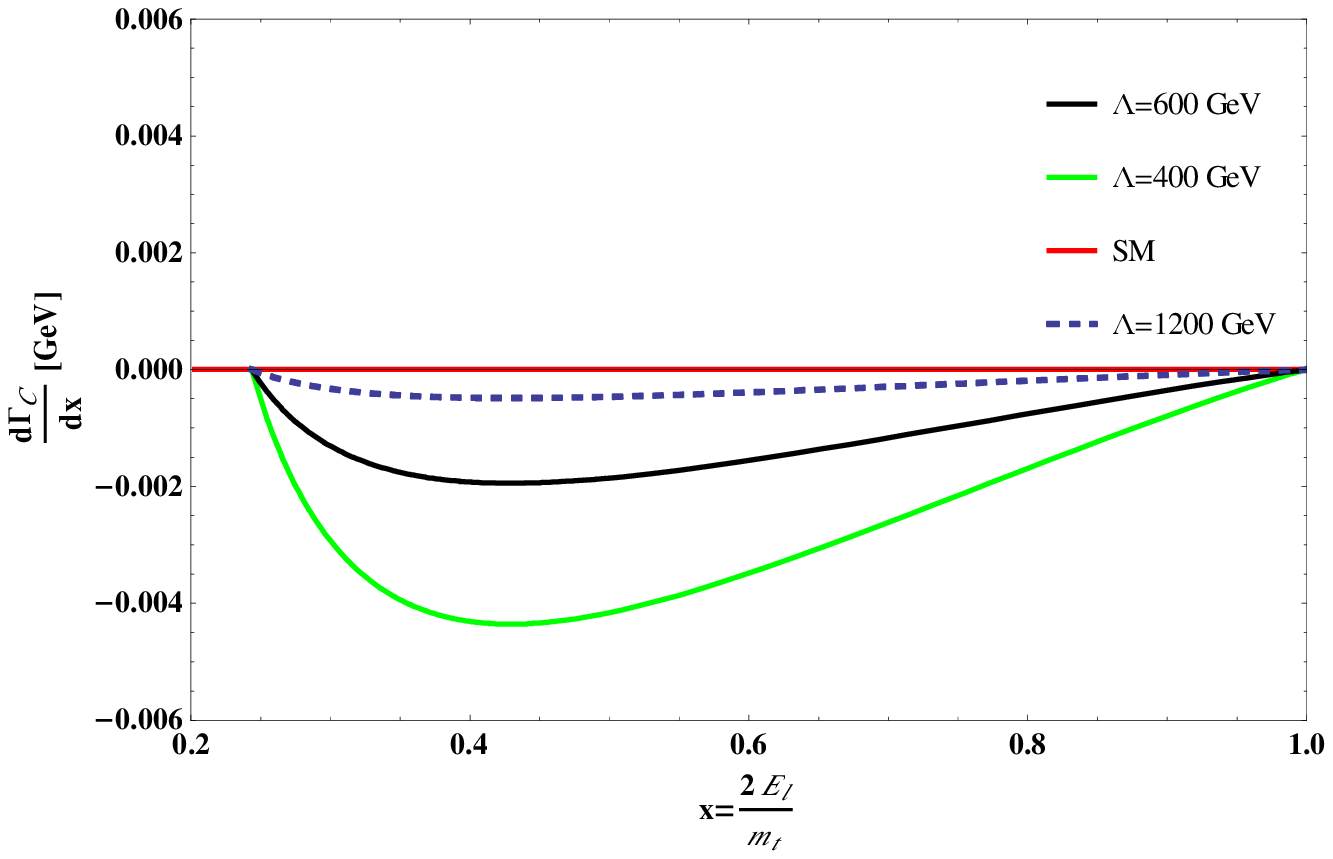}}   \caption{The azimuthal polarized differential rates w.r.t the charged lepton energy(charged lepton spectra).}
\label{fig.3} \end{figure}

Eq.(\ref{eq24}) shows the APDiR in NCST where it has been calculated up to the order of ${\theta ^2} \sim \frac{1}{{{\Lambda ^4}}}$. It is depicted in Fig \ref{fig.3}. In contrast to the LO of SM, the magnitude of the APDiR is non-zero in the LO of NCST. According to the Fig \ref{fig.3}, the APDiR decreases when NC scale increases and it vanishes when NC scale is around 1200 TeV and bigger.

\begin{figure}[t]
 \centerline{\includegraphics[width=0.5\textwidth]{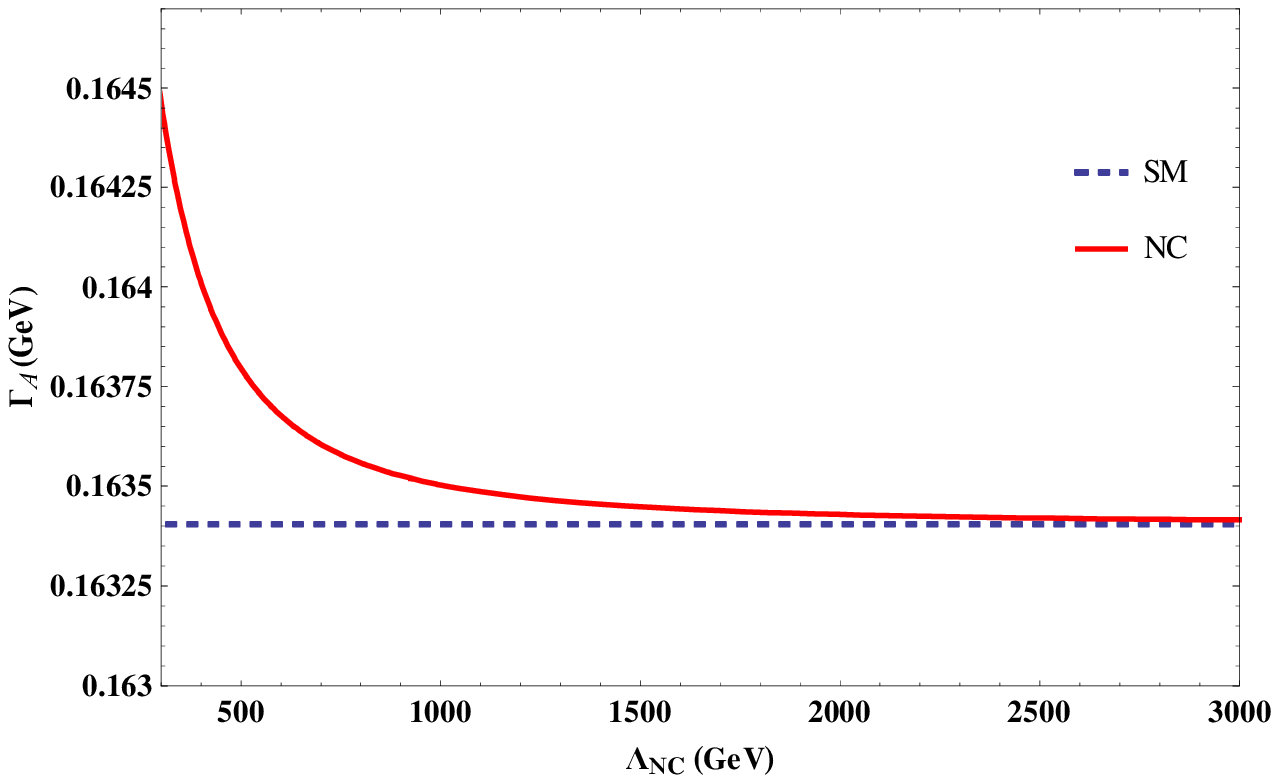}}   \caption{The unpolarized rates w.r.t NC scale in the logarithmic scale.}
\label{fig.4} \end{figure}

In sec.\ref{section.R3}, the integrated rates has been obtained up to the order of ${\theta ^2} \sim \frac{1}{{{\Lambda ^4}}}$, too. We have calculated the unpolarized rate up to the order of ${\theta ^2} \sim \frac{1}{{{\Lambda ^4}}}$ in Eq.(\ref{eq29}). Fig \ref{fig.4} shows the unpolarized top quark decay rate w.r.t the NC scale. This plot shows the deviation of the partial decay width from the SM expectations for $\Lambda  \le 2\;TeV$.

\begin{figure}[t]
 \centerline{\includegraphics[width=0.5\textwidth]{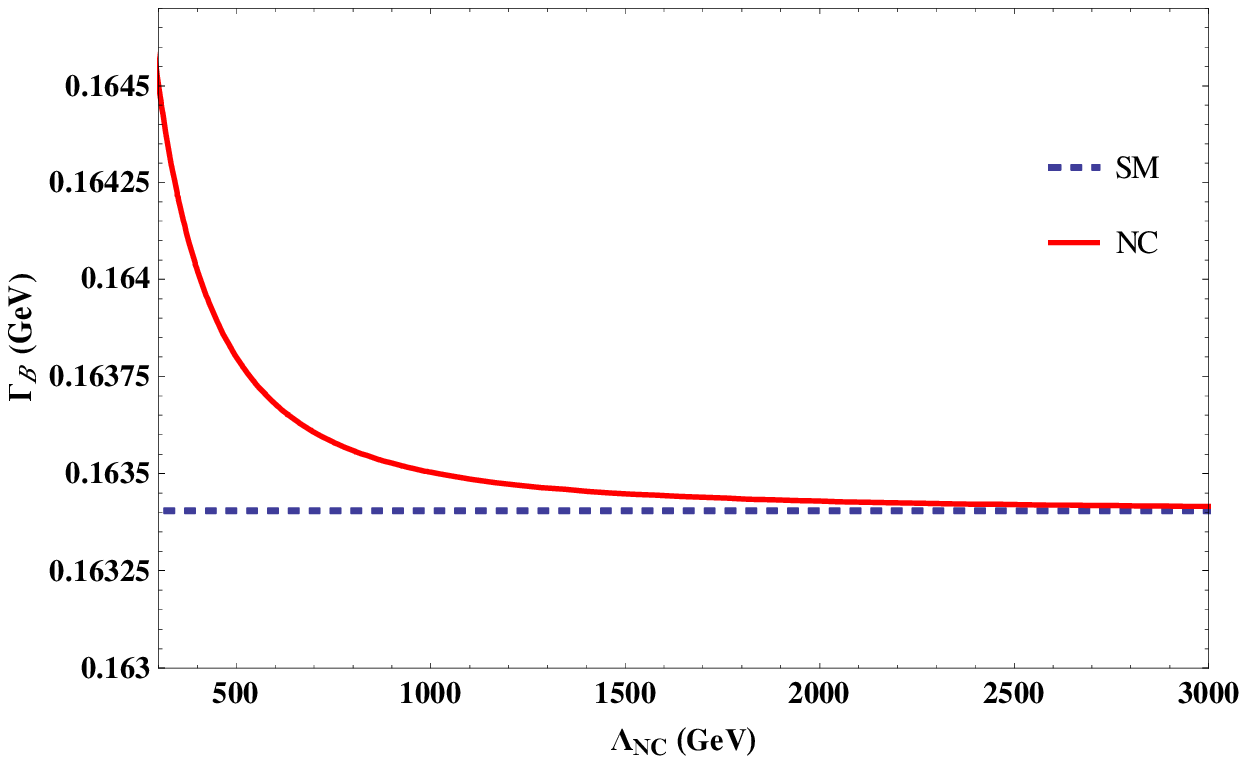}}   \caption{The polar polarized rates w.r.t NC scale in the logarithmic scale.}
 \label{fig.5} \end{figure}

The polar polarized rate up to the order of ${\theta ^2} \sim \frac{1}{{{\Lambda ^4}}}$ has been achieved in Eq.(\ref{eq30}). The dependence of PPDR to the NC scale can be seen in Fig \ref{fig.5}. The deviation of the partial decay width from the SM expectations is more obvious for $\Lambda  \le 2\;TeV$.

According to the first term in Eqs.(\ref{eq29}) and (\ref{eq30}), $\Gamma _A^{(0)}$ and $\Gamma _B^{(0)}$, it is clear that the UPDR and PPDR are equal in the LO of SM (see Ref.\cite{diss}), But they are not equal in NCST. It is noteworthy that two rates are equal up to the order of $\theta  \sim \frac{1}{{{\Lambda ^2}}}$ but are not equal in the order of ${\theta ^2} \sim \frac{1}{{{\Lambda ^4}}}$. The difference term is positive for the PPDR and negative for the UPDR, therefore the PPDR is bigger than the UPDR in the NCST ($\Gamma _B^{NC} > \Gamma _A^{NC}$), but it is insignificant and can be ignored.

\begin{figure}[t]
  \centerline{\includegraphics[width=0.5\textwidth]{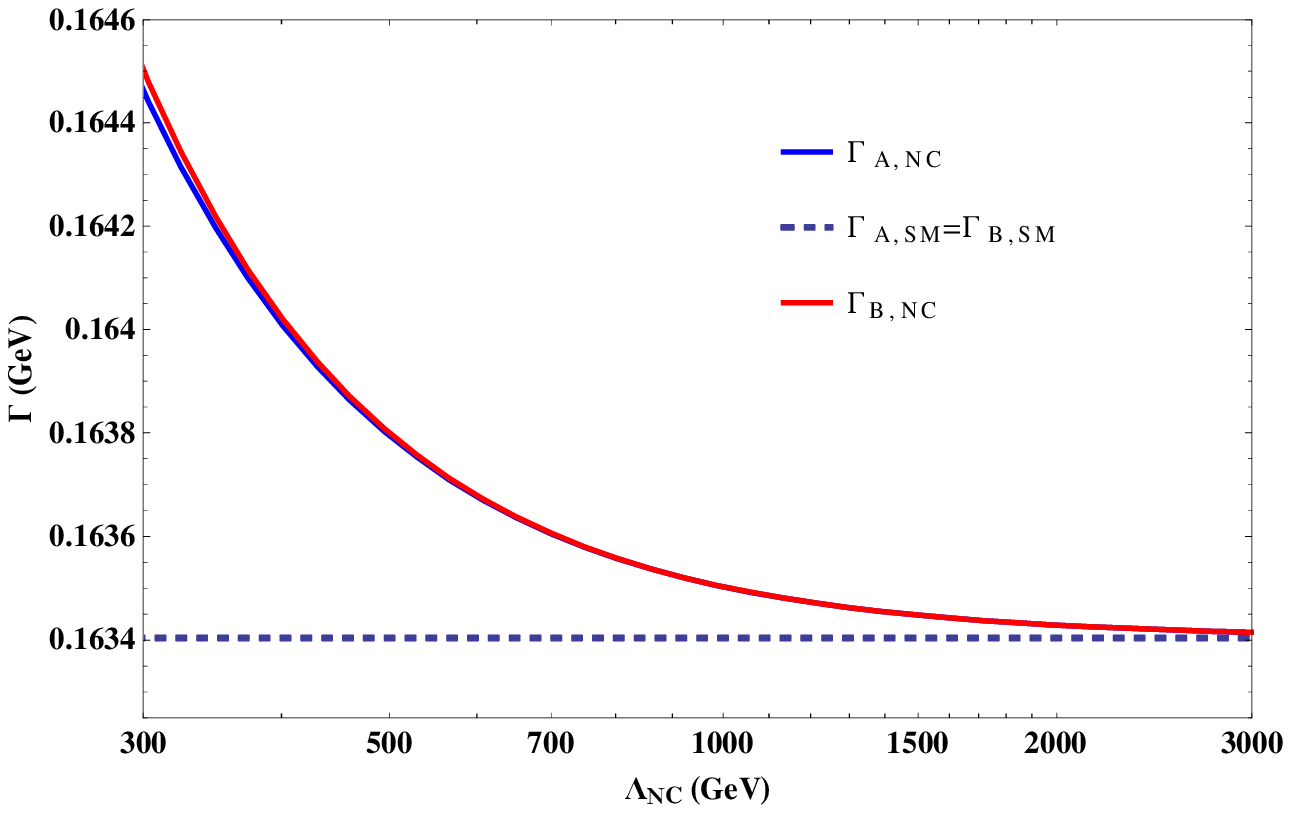}}   \caption{The difference of unpolarized and polar polarized rates w.r.t NC scale in the logarithmic scale.}
  \label{fig.6} \end{figure}

\begin{figure}[t]
 \centerline{\includegraphics[width=0.5\textwidth]{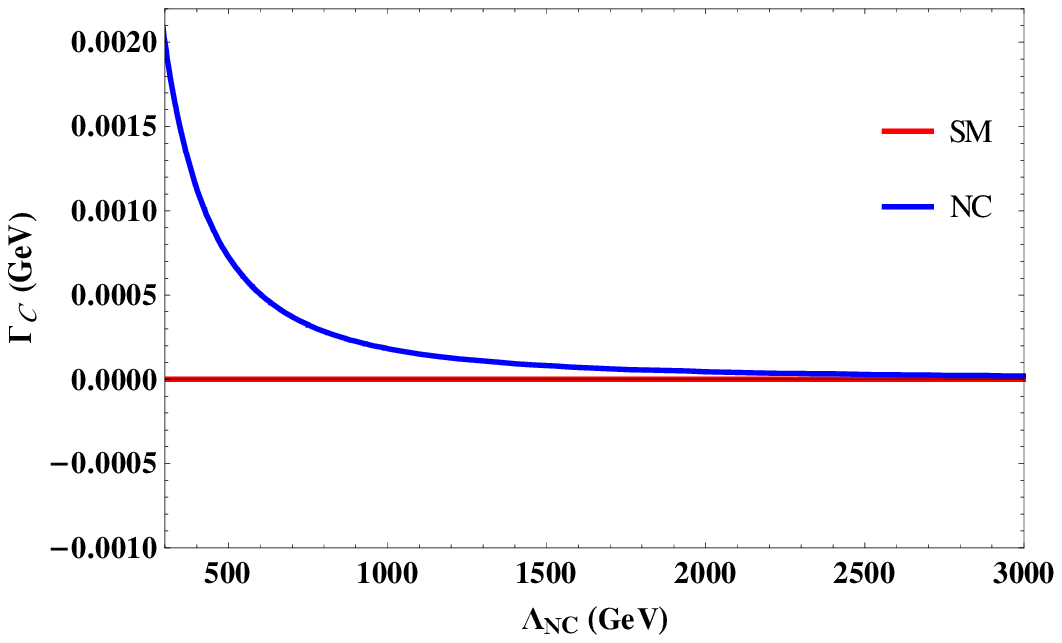}}   \caption{The azimuthal polarized rates w.r.t NC scale.}
 \label{fig.7} \end{figure}

The APDR has been achieved in Eq.(\ref{eq31}). It is just the contribution of NC space-time. The result of NC corrections on both leptonic and hadronic vertices for the azimuthal polarized rate has been shown in Fig \ref{fig.7}. According to the plot, the deviation of the partial width decay from SM expectations occurs when  $\Lambda  \le 1.5\;TeV$ and the results of NC tends toward the SM when $\Lambda  \ge 1.8\;TeV$. Of course, the reader should pay attention that as it is described at the end of the previous section, the specified limits in this section depend on the experimental values and its errors.

Now, we are going to describe some angular distribution results of Eq.(\ref{eq19}):

1) To obtain the total differential rate in the NCST, $\Gamma _{total}^{NC}$, we have used Eq.(\ref{eq19}). By getting the uniform angular distribution, i.e. $0 \le {\theta _P} \le \pi $ and $0 \le \phi  \le 2\pi $, and integrating on these intervals, we will have the following equality:

\begin{equation}\label{eq32}
\frac{{d\Gamma _{tota{\kern 1pt} l}^{NC}}}{{dx}} = \frac{{d\Gamma _A^{NC}}}{{dx}}
\end{equation}

as a result

\begin{equation}\label{eq33}
{\Gamma ^{NC}} = \Gamma _{total}^{NC} = \Gamma _A^{NC}
\end{equation}

These equations show that the uniform angular distribution leads to the equality of the total differential decay rate and the UPDiR and eventually to the equality of the total decay rate and the UPDR. Therefore Figs \ref{fig.1} and \ref{fig.4} display total  differential decay rate and total decay rate, too.

\begin{figure}[t]
 \centerline{\includegraphics[width=0.5\textwidth]{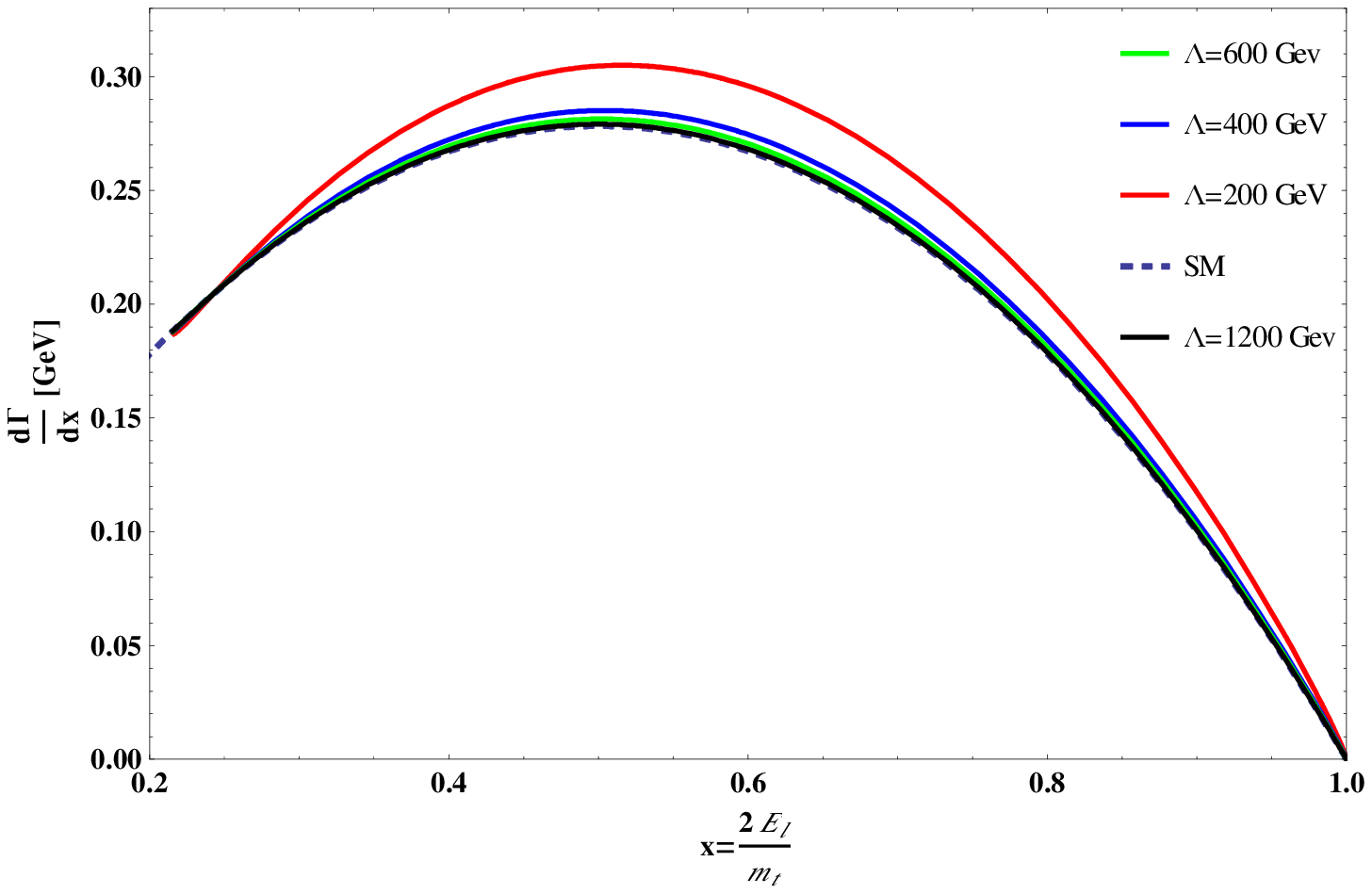}}   \caption{The energy spectra for the charged lepton in the polarized top quark semileptonic decay.}
\label{fig.8} \end{figure}

 Fig \ref{fig.8} shows the charged lepton energy spectra in the decay of unpolarized top quark by applying corrections on both leptonic and hadronic vertices.

\begin{figure}[t] 
 \centerline{\includegraphics[width=0.5\textwidth]{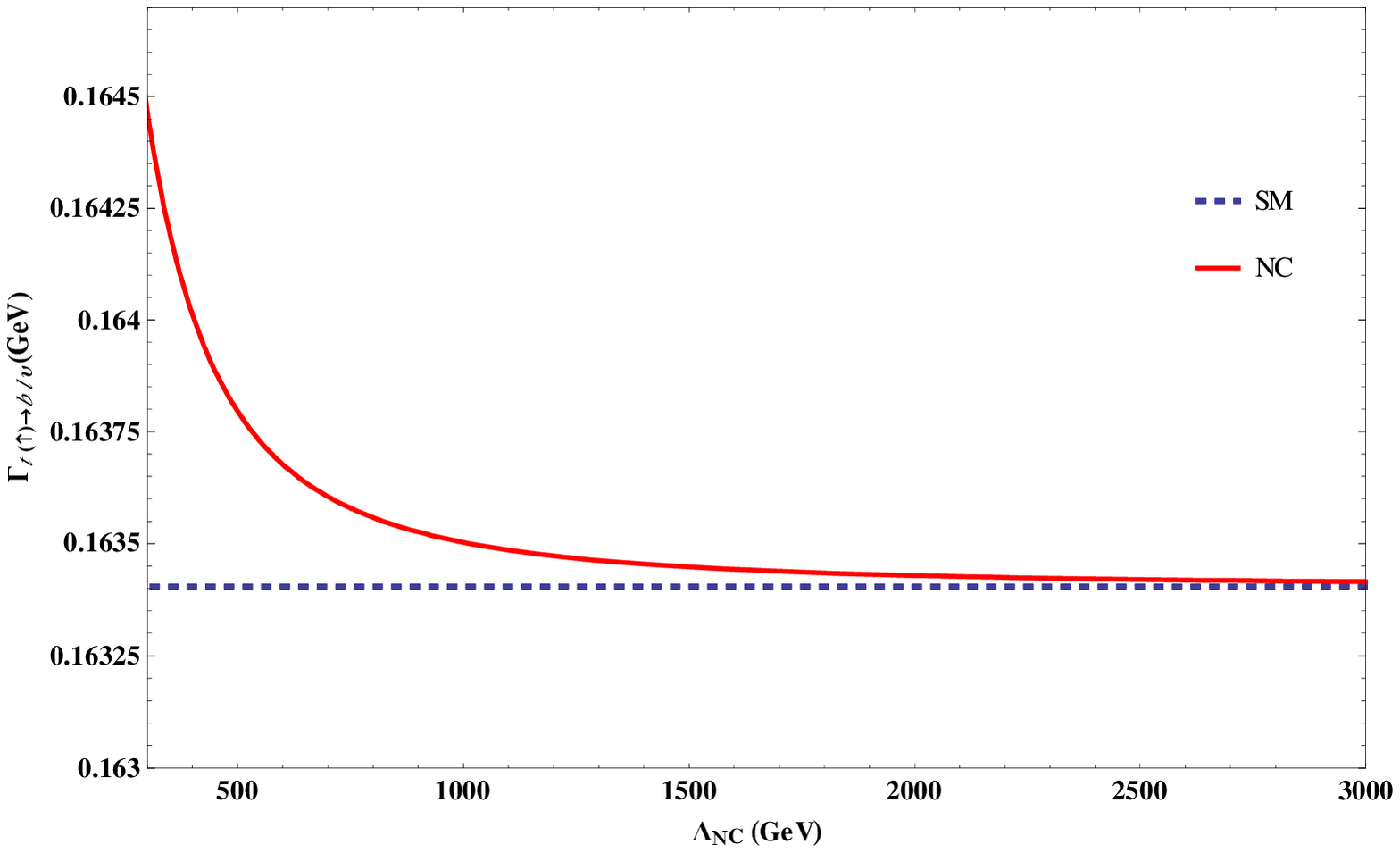}}   \caption{The total rate for $t( \uparrow ) \to b{\kern 1pt} {\ell ^ + }{\upsilon _\ell }$ decay w.r.t NC scale in the logarithmic scale.}
 \label{fig.9} \end{figure}

Fig \ref{fig.9} is expressing the total decay rate with regard to $\Lambda $ in the NCST by applying the corrected leptonic and hadronic vertices where the NC effects can be ignored for $\Lambda  \ge 1.5\;TeV$, whereas Fig.2 in Ref.\cite{ntblv} is expressing the total decay rate by applying the NC correction in the hadronic vertex and predicts the NC effects is invisible for $\Lambda  \ge 500\;GeV$.

\begin{figure}[t]   \centerline{\includegraphics[width=0.5\textwidth]{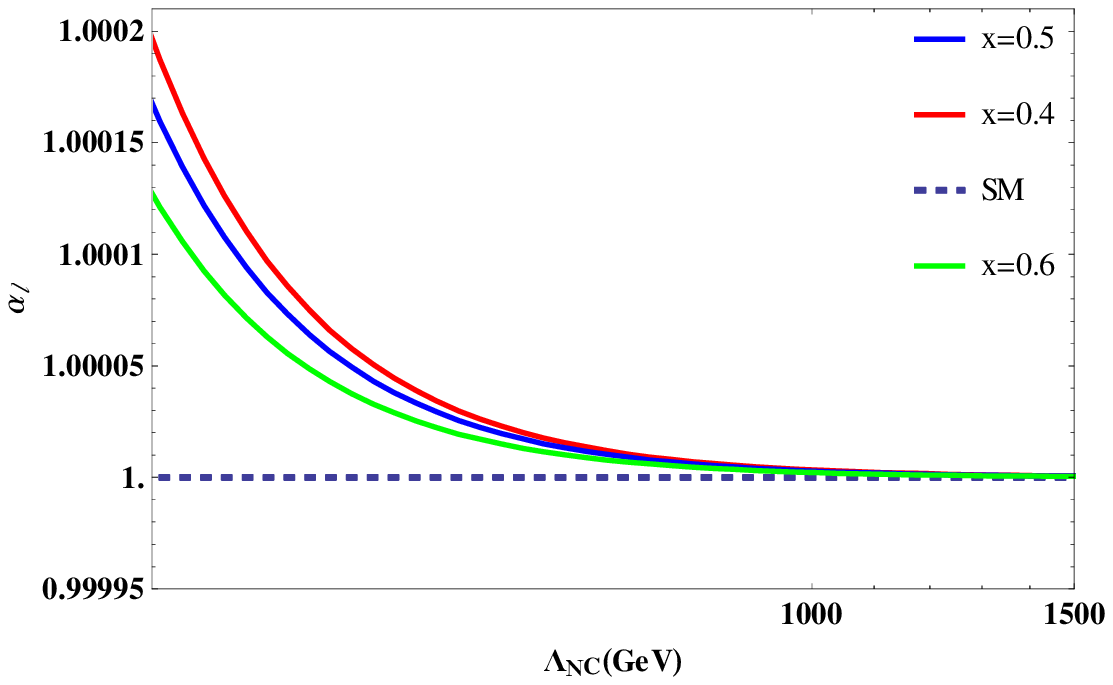}}   \caption{The spin analyzing power ${\alpha _\ell }(x)$ w.r.t NC scale in the logarithmic scale.}   \label{fig.10} \end{figure}

2) By inserting Eq.(\ref{eq32}) in Eq.(\ref{eq19}) and integrating on the $\phi$ azimuthal angle, we have
\begin{equation}\label{eq34}
\frac{{d{\Gamma ^{NC}}}}{{dxd\cos {\theta _P}}} = \frac{{d{\Gamma ^{NC}}}}{{dx}} \times \frac{1}{2}[1 + {\alpha _{{\kern 1pt} \ell }}(x)\cos {\theta _P}]
\end{equation}
and ${\alpha _\ell }(x)$, correction coefficient or spin analyzing power, is
\begin{equation}\label{eq35}
{\alpha _\ell }(x) = \frac{{\frac{{d\Gamma _B^{NC}}}{{dx}}}}{{\frac{{d\Gamma _A^{NC}}}{{dx}}}} = \frac{{{G_0}(x) + \frac{1}{{{\Lambda ^2}}}{G_1}(x) + \frac{1}{{{\Lambda ^4}}}{G_2}(x)}}{{{F_0}(x) + \frac{1}{{{\Lambda ^2}}}{F_1}(x) + \frac{1}{{{\Lambda ^4}}}{F_2}(x)}}
\end{equation}
where ${{F_0}(x)}$, ${{F_1}(x)}$, ${{F_2}(x)}$, ${{G_0}(x)}$, ${{G_1}(x)}$ and ${{G_2}(x)}$ have been defined in Eqs.(\ref{eq21}) and (\ref{eq23}). ${F_0}(x) = {G_0}(x)$, ${F_1}(x) = {G_1}(x)$, ${F_2}(x) \ne {G_2}(x)$ and ${F_2}(x) < {G_2}(x)$, therefore in NCST the spin analyzing power is larger than 1 (where it depends to x)whereas this coefficient equals to 1 in LO of SM (where it is independent of x or any other variable). Fig \ref{fig.10} indicates that the spin analyzing power decreases by increasing both the lepton energy and NC scale. The deviation from the SM is obvious for $\Lambda  \le 1\;TeV$ and the NC effect is ignorable for $\Lambda  > 1\;TeV$,

\begin{figure}[t]   \centerline{\includegraphics[width=0.5\textwidth]{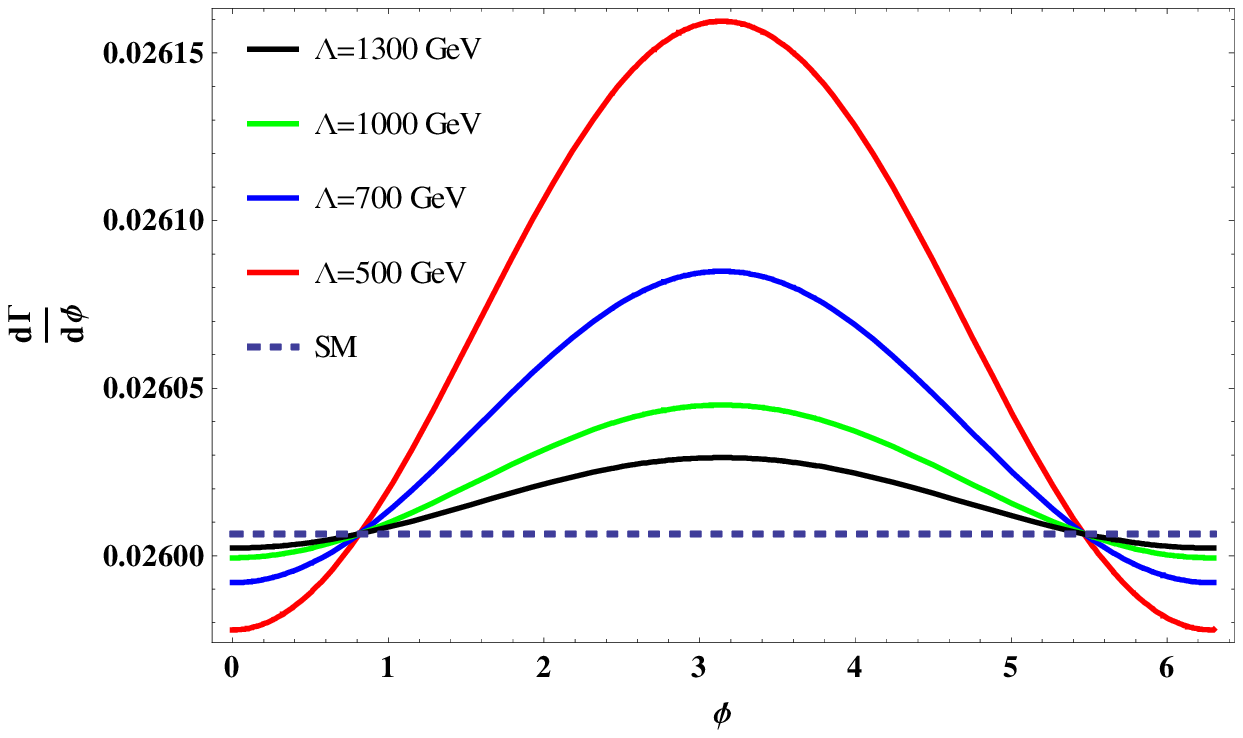}}   \caption{The dependence of total differential rate to the $\phi$ azimuthal angle and NC scale.}   \label{fig.11} \end{figure}

3) Eq.(\ref{eq36}) denotes the total  differential rate $\frac{{d{\Gamma ^{NC}}}}{{d\phi }}$ for $t( \uparrow ) \to b{\kern 1pt} {\ell ^ + }{\upsilon _\ell }$.
\begin{equation}\label{eq36}
\begin{array}{l}
\frac{{d{\Gamma ^{NC}}}}{{d\phi }} = \frac{1}{{4\pi }}[2\Gamma _A^{NC} - \frac{\pi }{2}\Gamma _C^{NC}\cos \phi ] = \frac{{1.11308}}{{4\pi }}[2(0.146803\\
\\
 - \frac{{288195}}{{{\Lambda ^4}}} + \frac{{88.9289}}{{{\Lambda ^2}}}) - \frac{\pi }{2}( - \frac{{211731}}{{{\Lambda ^4}}} + \frac{{164.045}}{{{\Lambda ^2}}})\cos \phi ]
\end{array}
\end{equation}
where $\Gamma _A^{NC}$ and $\Gamma _C^{NC}$ has been placed from Eqs.(\ref{eq29}) and (\ref{eq31}). It is depicted in Fig \ref{fig.11} at various values of $\Lambda $. It shows the total differential rate vs the $\phi $ azimuthal angle deviates from the SM. By increasing the NC scale, the deviation becomes small.

4) Eq.(\ref{eq19}) has been used to obtain the polar analyzing power or polar asymmetry in NCST, $\frac{{\Gamma _B^{NC}}}{{\Gamma _A^{NC}}}$. We have
\begin{equation}\label{eq37}
\frac{{d{\Gamma ^{NC}}}}{{d\cos {\theta _P}d\phi }} = \frac{{\Gamma _A^{NC}}}{{4\pi }}[1 + \frac{{\Gamma _B^{NC}}}{{\Gamma _A^{NC}}}\cos {\theta _P} + \frac{{\Gamma _C^{NC}}}{{\Gamma _A^{NC}}}\sin {\theta _P}\cos \phi ]
\end{equation}
The polar analyzing power in LO of SM is equal 1 and in NLO QCD is $0.998$ (see Ref.\cite{base}). The polar analyzing power in NCST and at various values of $\Lambda $ has been shown in Table \ref{tab.1}. According to Table \ref{tab.1}, the polar analyzing power in NCST is always bigger than 1 and by increasing the $\Lambda $, this value decreases and trends to 1.However it could be understand that the correction of NCST to the polar analyzing power is less than $0.003\%$ when $\Lambda=500 GeV $.

\begin{table}
\caption{The polar analyzing power, $\frac{{\Gamma _B^{NC}}}{{\Gamma _A^{NC}}}$, w.r.t NC scale.}
\label{tab.1}
\centering
\begin{tabular}{|l||r|r|r|c|}
\hline\hline
$\Lambda (GeV)$ & 200 & 500 & 900 & 1200\\
\hline
$\frac{{\Gamma _B^{NC}}}{{\Gamma _A^{NC}}}$ & 1.0012 & 1.00003 & 1.0000068 & 1.00000095\\
\hline
\end{tabular}
\end{table}

5) We obtain the NC corrections to the azimuthal rate, $\frac{{\Gamma _C^{{\kern 1pt} NC}}}{{\Gamma _A^{(0)}}}$, from Eq.(\ref{eq19}) and we have
\begin{equation}\label{eq38}
\begin{array}{l}
\frac{{d{\Gamma ^{NC}}}}{{d\cos {\theta _P}d\phi }} = \frac{{\Gamma _A^{(0)}}}{{4\pi }}[(1 + \frac{{\Gamma _A^{nc}}}{{\Gamma _A^{(0)}}}\% )\\
\\
 + (1 + \frac{{\Gamma _B^{nc}}}{{\Gamma _A^{(0)}}}\% )\cos {\theta _P} + (\frac{{\Gamma _C^{NC}}}{{\Gamma _A^{(0)}}}\% )\sin {\theta _P}\cos \phi ]
\end{array}
\end{equation}
According to Eq.(\ref{eq38}), $\frac{{\Gamma _C^{{\kern 1pt} NC}}}{{\Gamma _A^{(0)}}}$ is calculated at various values of $\Lambda $. It is shown in Table \ref{tab.2}.
\begin{center}
\begin{table}
\caption{ The NC corrections to the azimuthal rate, $\frac{{\Gamma _C^{{\kern 1pt} NC}}}{{\Gamma _A^{(0)}}}$, w.r.t NC scale.}
\label{tab.2}
\begin{tabular}{|c|c|c|}
\hline
$\Lambda (GeV)$ & $\Gamma _C^{NC}$ \\
\hline
  $ 200$  &  $27.03 \times {10^{ - 3}}\,\Gamma _A^{(0)}$ \\
\hline
  $500 $ &  $4.45 \times {10^{ - 3}}\,\Gamma _A^{(0)}$ \\
\hline
  $900$  &  $1.38 \times {10^{ - 3}}\,\Gamma _A^{(0)}$ \\
\hline
  $1200$  &  $0.76 \times {10^{ - 3}}\,\Gamma _A^{(0)}$ \\
\hline
\end{tabular}\label{tab.2}
\end{table}
\end{center}
By increasing the $\Lambda $, this ratio becomes smaller.  For $500\;GeV\le{\Lambda }\le 900\;GeV $ it is co-order with the value of NLO QCD ($\left| {\Gamma _C^{NLO}} \right| = 2.4 \times {10^{ - 3}}\Gamma _A^{(0)}$ in Ref.\cite{base,diss}). The NC effect on $\frac{{\Gamma _C^{{\kern 1pt} NC}}}{{\Gamma _A^{(0)}}}$  is equal to the NLO QCD when $\Lambda  \cong 680\;GeV$. It means that for this value of $\Lambda(= 680\;GeV) $, the correction of NC and NLO are the same.

\begin{equation}\label{eq39}
 \Lambda  \cong 680\,\,GeV,\quad \Gamma _C^{NC} = 2.4 \times {10^{ - 3}}\,\Gamma _A^{(0)}
\end{equation}

Note that the NLO contribution is negative and the NC contribution is positive , therefore the magnitude of NLO contribution is used.

\section{Conclusion}\label{section.R5}

We have calculated the effects of NCST on the semileptonic decay of polarized top quark. This decay has been investigated in the helicity system where the event plane defines the (x,z) plane and the z-axis is in the direction of the ${\ell ^ + }$ momentum. The calculations show that the unpolarized, polar polarized and azimuthal polarized  rates, the total differential rate, the spin analyzing power and the energy spectra of charged lepton receive some corrections from NCST.

The unpolarized and polar polarized (differential) rates receive small corrections from NCST, see Figs \ref{fig.4}-\ref{fig.5} (see Figs \ref{fig.1}-\ref{fig.2}), the azimuthal polarized (differential) rate that is zero in LO of SM, is given the non-zero value in NCST, see Fig \ref{fig.7}(see Fig \ref{fig.3}). It is equal to the contribution of NLO QCD when  $\Lambda \cong 680\;GeV$.  The total decay rate, Fig \ref{fig.9}, also get a bit correction. The spin analyzing power is sensitive to the NC scale. It is 1 in the LO of SM  whereas it is bigger than 1 when $\Lambda  \simeq 1\;TeV$ (Fig \ref{fig.10}). The deviation of energy spectra for charged lepton is depicted in Fig \ref{fig.8}. It shows that in the future colliders with access to high energies and the production of top quark in such energies, these may produce more charged lepton than the SM, because of NC effects.
The  dependence of total differential rate to the $\phi$ azimuthal angel has significant contribution in the NCST, Fig \ref{fig.11}.

We have used the case of ${\theta ^{\mu 0}} = 0$ to study the polarized top quark and we hope the case of ${\theta ^{\mu 0}} \ne 0$ will be reviewed in the future.

\section*{Acknowledgment}
The authors are grateful Dr Seyed Mohammad Moosavi-Nejad for useful discussions.

\end{document}